\documentclass[twoside,final,reqno]{birkart}

\usepackage{graphicx}
\usepackage{amssymb}

\begin{document}
\def\sgn{\mathop{\rm sgn}\nolimits}
\def\div{\mathop{\rm div}\nolimits}
\def\rot{\mathop{\rm rot}\nolimits}
\def\AU{{\rm AU}}
\def\const{\mathop{\rm const}\nolimits}
\def\minmod{\mathop{\rm min\,mod}\nolimits}
\def\tfrac#1#2{{\textstyle\frac{#1}{#2}}}
\def\fd#1#2{\frac{d#1}{d#2}}
\def\pd#1#2{\frac{\partial#1}{\partial#2}}
\def\brk#1{\ifx=#1=\hyphenation{}{}{}\fi}
\let\ds=\displaystyle
\def\diag{\mathop{\rm diag}\nolimits}
\def\det{\mathop{\rm det}\nolimits}
\def\sskip{\noalign{\kern 2pt}}
\mathchardef\0="7630

\setcounter{page}{1}

\title[Numerical Solution of Hyperbolic Systems]
{Mathematical Aspects of Numerical Solution\\
of Hyperbolic Systems}
\author[Kulikovskii, Pogorelov, and Semenov]{A.~G.~Kulikovskii$^1$,
N.~V.~Pogorelov$^2$, and A.~Yu.~Semenov$^3$}

\address{
$^1$Department of Mechanics,\br
Steklov Mathematical Institute,\br
Russian Academy of Sciences,\br
8, Gubkin St.,\br
117966 Moscow, GSP-1, Russia.}

\email{kulik@class.mi.ras.ru}

\address{
$^2$Institute for Problems in Mechanics,\br
Russian Academy of Sciences,\br
101 Vernadskii Avenue,\br
117526 Moscow, Russia.}

\email{pgrlv@ipmnet.ru}

\address{
$^2$VKIV Department,\br
General Physics Institute,\br
Russian Academy of Sciences,\br
38 Vavilov St., 117942 Moscow, Russia.}

\email{say@lpl.gpi.ru}

\begin{abstract}
A number of physical phenomena are described by nonlinear hyperbolic
equations. Presence of discontinuous solutions motivates the necessity
of development of reliable numerical methods based on the fundamental
mathematical properties of hyperbolic systems. Construction of such
methods for systems more complicated than the Euler  gas dynamic
equations requires the investigation of existence and uniqueness
of  the self-similar solutions to be used in the development of
discontinuity-capturing high-resolution numerical methods. This frequently
necessitates the study of the behavior of discontinuities  under vanishing
viscosity and dispersion. We discuss these problems in the application to
the magnetohydrodynamic equations, nonlinear waves in elastic media,
and electromagnetic wave propagation in magnetics.
\end{abstract}

\maketitle

\section{Introduction}
In this paper we discuss the mathematical aspects of the problems
originating in the
solution of nonlinear systems of hyperbolic partial differential equations.
These equations describe a large variety of physical phenomena, such as,
gasdynamics, magnetohydrodynamics (MHD), shallow water equations,
elasticity equations, etc. Being nonlinear, these systems usually require
numerical methods for their solution. Presence of discontinuous solutions
motivates the necessity of the development of reliable numerical methods
based on the fundamental mathematical properties of hyperbolic systems.
Although such methods are rather well developed for the Euler gasdynamic
equations in the conservation law form, their extension to more complicated
hyperbolic systems is not straightforward. It requires a mathematical
justification of the solution uniqueness, a formulation of the selection
principles for relevant solutions, and, finally, an investigation of their
physical validity. Most of high-resolution methods for gasdynamic equations
use the exact or some of the approximate self-similar Riemann problem
solutions to determine fluxes through the computational cell surfaces.
Similar methods are expected to be developed for various types of
hyperbolic systems.
In this case we must construct the elementary self-similar solution
using only admissible discontinuities (entropy consistent, evolutionary,
etc.).
Basically the choice of the solution must be
made on the basis of the structure of the solution
of the extended problem \cite {Kul90}.
All mentioned above makes very important the study of discontinuous
solutions behavior under vanishing viscosity and dispersion to create a
proper background for the development of high-resolution numerical methods for
hyperbolic systems
more complicated than the Euler equations of gasdynamics.
We discuss several analytical and numerical solutions in the mentioned
fields which illustrate the complexity of the selection problem and outline
the methods of its solution.

\section{High-resolution methods for MHD equations}
TVD upwind and symmetric differencing schemes have recently become very
efficient tool for solving complex multi-shocked gasdynamic flows. This is
due to their robustness for strong shock wave calculations.
The extension of these schemes to the
equations of the ideal magnetohydrodynamics is not simple.
First, the exact solution  \cite{Kul65} of the MHD Riemann problem is
too multivariant to be used in regular calculations.  Second, several
different approximate solvers \cite{Aslan}, \cite{Barmin96}, \cite{Brio},
\cite{Dai}, \cite{Hanawa}, \cite{Pog95}, and \cite{Powell}
applied to MHD equations are now at the stage of
investigation and comparison. This investigation requires i)
determination of a proper slope limiting method in the parameter
interpolation procedure necessary to obtain nonoscillatory schemes
of the order of accuracy higher than one; ii) development of an
efficient entropy correction method necessary to exclude rarefaction
shocks; and, finally, iii) solution of the problem of excluding the origin
of nonevolutionary solutions in ideal MHD calculations.

The system of governing equations for a MHD flow of an ideal,
infinitely conducting, perfect plasma in the  Cartesian coordinate
system $x$,~$y$,~$z$ with the use of the conventional notations reads (one
fluid approximation):
\begin{equation}
{\partial {\bf U}\over \partial t}
  + {\partial {\bf E}\over \partial x} + {\partial {\bf F}\over \partial y}
  + {\partial {\bf G}\over \partial z} + {\bf H}_{\rm div} =
  \mathchar"0630\, ,
\end{equation}

\noindent
where $\bf U$ is the vector of conservative variables and $\bf E$, $\bf F$,
and $\bf G$ are the flux vectors.

We introduced here the source term ${\bf H}_{\rm div}$ in the form
\[
  {\bf H}_{\rm div}= \div {\bf B} \times \left(
  0,\,{B_x \over 4\pi},\,{B_y \over 4\pi},\,{B_z \over 4\pi},\,
  {{\bf v}\cdot {\bf B}\over 4\pi},\,u,\,v,\,w\right)^{\rm T}.
\]

This form of the system can be used to satisfy the divergence-free
condition by convecting away the magnetic charge from the computational
region \cite{Powell}. Otherwise, any other well-known method can be used
to eliminate the magnetic charge.

To determine a numerical flux
$ \bar{\bf E} = n_1 {\bf E} + n_2{\bf F} + n_3{\bf G}$
normal to the computational cell boundary
(${\bf n}= (n_{1},n_{2},n_{3})$ is a unit outward vector normal
to the cell surface) one can use the formulas based on the
solution of the linearized problem
\begin{equation}
\bar {\bf E}({\bf U}^R,{\bf U}^L)={1\over 2}
\left[{\bf E}({\bf U}^L) + {\bf E}({\bf U}^R) -S|\Lambda|S^{-1}({\bf U}^R -
{\bf U}^L)\right].
\end{equation}

Here $S(\bar{\bf U})$ and $S^{-1}(\bar{\bf U})$ are the matrices formed by
the right and by the left eigenvectors, respectively, of the frozen
Jacobian matrix
\[
  \bar J = {\partial \bar{\bf E}({\bf U}^L, {\bf U}^R)\over \partial {\bf
   U}}.
\]

The matrix $|\Lambda|$ is a diagonal matrix consisting of the frozen
Jacobian matrix eigenvalue moduli.
The superscripts $R$ and $L$ denote the values at the right- and at the
left-hand side of the cell boundary.

In pure gas dynamics the uniform average vector $\bar{\bf U}({\bf U}^L,
{\bf U}^R)$ can be constructed in such a way that the conservation
relations on shocks are exactly satisfied.  The important peculiarity of
the latter method is that, although it gives the solution of the linearized
problem, the exact satisfaction of the Rankine--Hugoniot relations on
shocks provides their more adequate and sharp resolution \cite{Roe}.

In \cite{Brio} the MHD numerical Jacobian matrix was used frozen at the
point obtained by the arithmetic average between ${\bf U}^R$ and ${\bf
U}^L$.  Thus, it could not guarantee the exact satisfaction of the
conservation condition at the jump. This approach belongs to the general
conservative Courant--Isaacson--Rees family.

The reason of the former averaging for the
MHD equations is explained by the fact that there is no single
averaging procedure to find a frozen Jacobian matrix of the system.
Another linearization approach is used in  \cite{Aslan}, \cite{Hanawa},
\cite{Pog95} in which  the linearized Jacobian matrix is not a function of
a single averaged set of variables, but depends in a complicated way on the
variables on the right- and on the left-hand side of the computational cell
surface. In \cite{Pog96} and \cite{Pog97}  this procedure was shown to be
nonunique.  A multiparametric family of linearized MHD approximate Riemann
problem solutions was presented that assured an exact satisfaction of the
conservation relations on discontinuities. A proper choice of parameters is
necessary to avoid physically inconsistent solutions.

It is widely known that the interpolation method used to determine parameter
values at the cell surfaces can greatly improve the quality of
numerical results. Let us introduce the mesh function ${\bf U}^n_i =
{\bf U}(n\Delta t, i\Delta x)$. We shall
use as an example the following interpolation approach \cite{Barmin96}:
\begin{eqnarray*}
 & &{\bf U}^R_{i+1/2} = {\bf U}^{n}_{i+1} - {\textstyle{1 \over 4}} [(1 -
 \eta) \tilde{\Delta}_{i+3/2} + (1 + \eta)
 \tilde{\tilde{\Delta}}_{i+1/2}],\\
 & &{\bf U}^L_{i+1/2} = {\bf U}^{n}_{i}
 + {\textstyle {1 \over 4}} [(1 - \eta) \tilde{\tilde{\Delta}}_{i-1/2} + (1 
+ \eta) \tilde{\Delta}_{i+1/2}],\\ & &\tilde{\Delta}_{i+1/2} = 
 \minmod(\Delta_{i+1/2}, \omega \Delta_{i-1/2}), \ \
 \tilde{\tilde{\Delta}}_{i+1/2} =
 \minmod(\Delta_{i+1/2}, \omega \Delta_{i+3/2}),\\
 & & \minmod(x,y) = \sgn (x) \times \max \{0, \min [|x|, y \sgn (x)]\}.
     \phantom{\tilde{\tilde{\Delta}}_{i+1/2}}
\end{eqnarray*}

The choice $\eta = -1$ and $\omega =1$ gives a popular ``$\minmod$'' method
which usually eliminates spurious oscillations near discontinuities.
On the other hand, the application of more compressive slope limiters
is widely accepted in gas dynamics  for finer
resolution of contact discontinuities. In MHD the same is important for
rotational (Alfv\'enic) discontinuities.
We can suppose the choice $\eta = {1 \over 3}$ (the third-order
upwind-biased interpolation) and $\omega = 2$ to give better results.
Consider as an example the MHD Riemann problem with the following initial
conditions ($\rho$, $p$, $u$, $v$, $w$, $B_{y}/\sqrt{4
\pi}$,~$B_{z}/\sqrt{4 \pi}) = (0.18405$, 0.3541, 3.8964, 0.5361, 2.4866,
 2.394, 1.197) for $x < 0.5$ and (0.1,~0.1, $-5.5$, 0, 0, 2,~1) for $x >
0.5$ with $B_x \equiv 4$ and the specific heat ratio $\gamma=1.4$.
The solution of this problem contains all types of MHD shocks
propagating through the both halves of the computational region
separated by the contact discontinuity \cite{Barmin96}. The
distribution of the $B_z$ component of the magnetic field vector is
presented for $t = 0.15$ (400 cells were taken between 0 and~1).  One can
clearly see in Fig.~1 the numerical noise attendant in this case in the
vicinity of strong shocks, which is similar to that being suppressed by the
artificial viscosity in \cite{Dai}.

\begin{figure}[tbp]
\begin{center}
\includegraphics[width=11cm,height=9cm,angle=0]{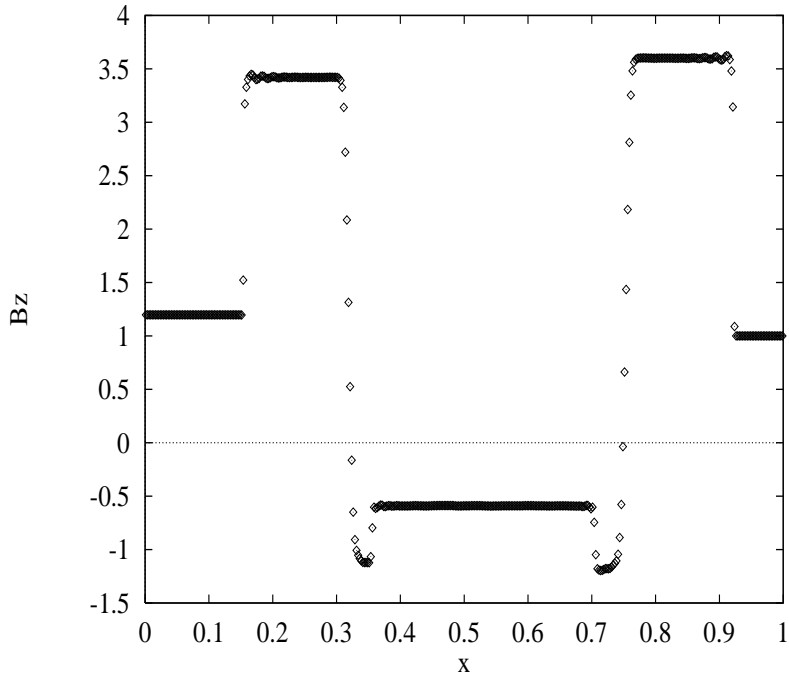}
\caption{$B_z$ distribution for $\eta= 1/3$}
\end{center}
\end{figure}

\begin{figure}[tbp]
\begin{center}
\includegraphics[width=11cm,height=9cm,angle=0]{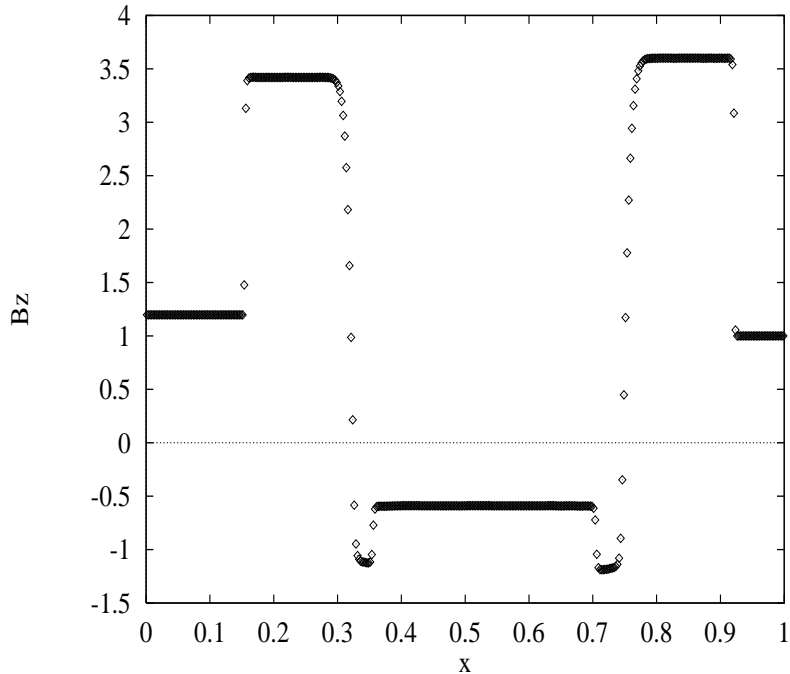}
\caption{$B_z$ distribution for $\eta= 1$}
\end{center}
\end{figure}

If one decides, however, to choose $\eta =1$ (three-point central
differencing scheme) these oscillations disappear (Fig.~2). This result
seems to have no analogue in purely gas dynamic calculations and speaks in
favor of the application of the central schemes rather than upwind ones.

Talking about the Roe-type solvers for MHD, it is worth noting that
application of the one-dimensional solver for multidimensional calculations
even with the use of Powell's technique \cite{Powell} cannot guarantee
the exact satisfaction of the Rankine--Hugoniot conditions due to the
presence of the artificial source term. This makes
questionable the necessity of utilization of the very complicated and time
consuming algorithms like \cite{Pog96}. Our experience based on the
solution of several one-dimensional and two-dimensional test problems
shows that the scheme \cite{Brio} gives essentially the same accuracy.

Another important thing of discussion is that certain initial
and boundary value problems can be solved nonuniquely using different
shocks or combination of shocks, whereas physically one would expect only
unique solutions. The situation differs from that in pure gasdynamics where
all entropy increasing solutions are evolutionary and physically relevant.
Contrary to pure gasdynamics, in the MHD case the condition of the entropy
increase is necessary but not sufficient. Only slow and fast MHD shocks are
evolutionary, while intermediate shocks are to be excluded.
Although in the ideal MHD a nonevolutionary shock decomposition into
evolutionary ones occurs instantaneously under action of an infinitesimal
perturbation, this decomposition in the presence of numerical
dissipation can require some time which depends on the numerical scheme
and grid resolution~\cite{Barmin96}. From this viewpoint the question
arises whether the schemes are applicable which use one-dimensional
coplanar Riemann problem solution to determine numerical fluxes at cell
interfaces. Among the schemes mentioned above only \cite{Dai} effectively
incorporates Alfv\'enic shocks into the flux determination procedure.
On the other hand, it is well known that boundary conditions for rotational
perturbations in MHD split from the full set of boundary conditions on the
shock.  That is why, the evolutionarity properties for this kind of
perturbations must be checked separately. Does this mean that
axisymmetric problems must be solved as three-dimensional or some
automatic algorithm can be constructed allowing one to introduce
rotational discontinuities in the framework of the two-dimensional
statement of the problem?  The answer to this question is still open. One
should also admit that the destruction time of the latter kind of
nonevolutionary waves can be rather long.  Although it is clear that
shocks unstable to tangential perturbations can be stable in dissipative
MHD, their behavior in the case of uncontrollable numerical dissipation is
hardly predictable.

In the context of this section we would like to indicate a very simple
numerical algorithm which was proposed by N.~Pogorelov in \cite{Barmin96}.
In this approach instead of Eq.~(2) we use the formula
\begin{eqnarray}
& &{\bar {\bf E}}_{i+1/2, n} = \frac {1}{2} \left[{\bar {\bf E}}\left({\bf
  U}^R_{i+1/2} \right) + {\bar {\bf E}}\left({\bf U}^L_{i+1/2}\right)
  +{\bf \Phi}_{i+1/2}\right]\, , \nonumber \\
& &  \\
& &{\bf \Phi}_{i+1/2} = -
  {\bf {\hat R}}_{i+1/2} \left({\bf U}^R_{i+1/2} - {\bf U}^L_{i+1/2}
   \right), \nonumber
\end{eqnarray}
where ${\bf {\hat R}}_{i+1/2}$ is the diagonal matrix with the same
elements on its diagonal equal to the spectral radius r (the maximum of
eigenvalue magnitudes) of the Jacobian matrix
${\partial {\bar {\bf E}} \over \partial {\bf U}}$.

This scheme can be treated as a second-order nonoscillatory extension of
the
Lax--Friedrichs method. It is extremely robust and automatically satisfies
the entropy condition, thus allowing one to avoid the application of
artificial entropy fix procedures. Several complicated axisymmetric
and three-dimensional physical problems (see \cite{PogSem} and
\cite{PoMat}) were successfully solved using this scheme.

\section{The solution selection problem}
We shall describe here several physical problems for which the Riemann
problem has
nonunique solutions. This nonuniqueness is merely caused by the
formulation of the problem,
since in the classical physics the future can be predicted uniquely by initial
conditions. The oversimplification is connected with the following two
circumstances.

Alongside with the hyperbolic system of equations presupposing the presence of
discontinuities in its solutions, there exist as a rule a more
complete system which
takes into account such processes as viscosity, heat conduction, diffusion,
finite electric conductivity, etc. This system transforms into the
hyperbolic one for large-scale phenomena if certain small terms are
neglected. The complete system usually
has only continuous solutions which can be supposed  unique. The
mentioned system,
however, has no self-similar solutions depending only on $x/t$.
Self-similar
solutions  of this kind can originate as the asymptotics of solutions of
appropriate
problems for  the complete system for $t \to \infty$. Let us consider as
well-selected
the self-similar  solution which is the asymptotics of some definite
problem for the
complete system.  The suggested selection principle clearly depends on
the physical processes
which are not taken  into account in the simplified hyperbolic system. In
the theoretical
study it also depends on the selection of the complete system itself.

Note that  not only solutions corresponding to
discontinuous initial conditions but also solutions with arbitrarily (and
not necessarily monotonically) smeared discontinuities in initial
conditions  have the same asymptotics $x/t$. The particular
specification of this smearing can affect the establishment of one or
another asymptotics.
This is the second circumstance which influence the correct choice of  the
self-similar solution.

The first step in the application of the complete system of equations is
usually represented by selection among all discontinuities satisfying
the conservation laws
those ``admissible''  discontinuities which have a corresponding solution
of the discontinuity structure problem described by the complete system.
As a discontinuity structure, we imply the solution  which represents
a continuous variation of values corresponding to the jump, this
solution being usually considered one-dimensional and
stationary. Note that the latter limitations are not always satisfied for
realistic discontinuities. The requirement of existence of this structure
provides the entropy nondecrease condition and, if the conservation laws
are not sufficient for the discontinuity to be evolutionary, is used to
obtain new additional boundary conditions whose satisfaction can make the
discontinuity evolutionary. The latter statement was
proved for the case of the stationary one-dimensional structure
\cite{Kul68} and also for
the case of periodicity with respect to time and space variables along the
front (cell
structure) \cite{Kul90}.

In a number of cases the requirement of admissibility of all discontinuities
turns out to be sufficient for the solution uniqueness. As an
example we can mention the problem of motion of the gas with a complicated
equation of state \cite{Galin} where viscosity and heat conductivity are
taken into account in the discontinuity structure. On the other hand, there
exist problems for which the requirement of admissibility is insufficient
for the uniqueness of the self-similar solution. One of the well-known
examples is the decay of an arbitrary jump in a combustible gas mixture.
There exist mixtures for which both detonation and slow combustion are
possible depending on the ignition method, that is, on the smearing method
of the initial discontinuity (for the detonation to be realized we need a
nonmonotonic smearing with the energy excess).
Thus, we encounter a physical unremovable nonuniqueness of the Riemann
problem solution. New problems with a similar nonremovable nonuniqueness
have lately been discovered.

1. {\em The theory of elasticity with viscosity taken into account in the
study of small-scale phenomena.} Nonuniqueness of self-similar problems
was found in the investigation of nonlinear
quasi-transverse small-amplitude waves in a weakly anisotropic elastic
media \cite{KulSvesh} , which occurs in the general case of one uses only
admissible discontinuities. Numerical experiments \cite{Chug90},
\cite{KulSvesh} with viscosity taken into account showed that, depending on
the details of the problem statement  which are not taken into account in
the simplified (hyperbolic) model, all available  self-similar solutions
can be realized as an asymptotics of the solution for $t \to \infty$.
However, under monotonic smearing of the initial conditions the solution
always follows the asymptotics of certain definite type.

2. {\em Nonlinear electromagnetic waves in magnetics} \cite{Gvoz97}. In this
case the equations and the jump relations do not differ in the large-scale
approximation from those describing quasi-transverse elastic waves.
However, the structure of electromagnetic shock waves is connected with
completely different mechanisms which create the dispersion of short waves.
The variety of admissible discontinuities turns out for this reason
completely different from the corresponding variety in the theory of elasticity.
In particular, a set of separate points corresponding to discontinuities
with one additional condition lie on the shock adiabatic curve. The number
of these points is determined by the ratio between the dispersion and
viscosity effects inside the structure and can be as
large as several tens for real  magnetics, thus leading to the multiple
nonremovable nonuniqueness of self-similar problems.

3. As shown in \cite{Gvoz98}, the problem on {\em the longitudinal wave
propagation through the rod} has properties similar to those described in
the previous example if the derivative of the rod tension with  respect to
its stress in nonmonotone and has at least two extremums (a minimum and a
maximum). Earlier in \cite{Kul84} a mathematical example with exactly the
same jump behavior was considered. It was based on the first order
equation taking into account the dispersion and dissipation under wave
propagation .  The shock behavior and connected with this multiple
nonuniqueness of the solutions of the self-similar problems was shown in
\cite{Gvoz98} to be quite usual if the shock
structure is determined by the equations  with dispersion which causes the
oscillation of the parameters of the medium inside the discontinuity
structure.

4. Self-similar  solutions can frequently be nonunique or even nonexistent
if {\em one of the discontinuities is represented by a phase transition
front}.  We shall mention here only the problems dealing with ionization
and recombination fronts. A systematic survey of this subject can be found
in \cite{Barmin71}. We briefly discuss only solutions describing the motion
of a gas in a magnetic field where the fronts exist of  the gas electrical
conductivity switch-on \cite{Barm75}. The investigation showed that the
self-similar problem can have one self-similar
solution, or several ones, or none of them, depending on the
choice of parameters .  Numerical experiments \cite{Barm86} undertaken in
the assumption that the only dissipative mechanism is represented by the
finite gas conductivity showed that if self-similar solutions exist then
the numerical solution asymptotically approaches one of them as $t$
increases.  If a self-similar solution does not exist, alternating layers
of conductive and nonconductive gas originate in calculations and the
number of these layers increases in time.

{\bf Acknowledgment.} This work was supported, in part, by the Russian
Foundation for Basic Research Grants No.~96-01-00991 (A.~G.~K.) and
No.~98-01-00352 (N.~V.~P. and A.~Yu.~S.). A~.G.~K. was also supported
by the INTAS--RFBR Grant No.~95-0435.

\end{document}